\documentclass[prd,aps,showpacs,floats,letterpaper,floatfix,
groupedaddress,superscriptaddress
,nofootinbib
,eqsecnum
]{revtex4}
\usepackage{amsmath}
\usepackage{amsfonts}
\usepackage{bm}
\usepackage{graphicx}

\newcommand\be{\begin{equation}}
\newcommand\ee{\end{equation}}
\newcommand\nn{\notag \\ }

\newcommand{\tp}{\tilde{\psi}}
\newcommand{\tm}{\tilde{m}}
\newcommand{\tl}{\tilde{L}}
\newcommand{\tls}{\tilde{L}^2}
\newcommand{\te}{\tilde{\epsilon}}
\allowdisplaybreaks

\def\prd{{Phys.\ Rev.\ D }}
\def\cqg{{Class.\ Quantum Grav.\ }}

\begin{document}

\title{Dark matter distributions around massive black holes: A general relativistic analysis}

\author{
Laleh Sadeghian} \email{lsadeghian@wustl.edu}
\affiliation{McDonnell Center for the Space Sciences, Department of
Physics, Washington University, St.  Louis, Missouri 63130, USA}
\affiliation{Department of Physics, University of Florida, Gainesville, Florida 32611, USA}

\author{
Francesc Ferrer} \email{ferrer@physics.wustl.edu}
\affiliation{McDonnell Center for the Space Sciences, Department of
Physics, Washington University, St.  Louis, Missouri 63130, USA}

\author{
Clifford M.~Will} \email{cmw@physics.ufl.edu}
\affiliation{Department of Physics, University of Florida, Gainesville, Florida 32611, USA}
\affiliation{GReCO, Institut d'Astrophysique de Paris, CNRS,\\ 
Universit\'e Pierre et Marie Curie, 98 bis Bd. Arago, 75014 Paris, France}

\date{\today}

\begin{abstract}

The cold dark matter at the center of a galaxy will be redistributed by the presence of a massive black hole.
The redistribution may be determined using an approach pioneered by Gondolo and Silk: begin with a model distribution function for the dark matter,
and ``grow'' the black hole adiabatically, holding the adiabatic invariants of the motion constant. Unlike
the approach of Gondolo and Silk, which adopted Newtonian theory together with ad hoc correction factors to mimic
general relativistic effects, we carry out the calculation fully relativistically, using the exact Schwarzschild
geometry of the black hole.  We find that the density of dark matter generically vanishes at $r=2R_{\rm S}$, not $4R_{\rm S}$ as found by Gondolo and Silk, where $R_{\rm S}$ is the Schwarzschild radius, and that the spike very close to the black hole reaches significantly higher densities. 
We apply the 
relativistic adiabatic growth framework to obtain the final dark matter
density for both cored and cusped initial distributions.
Besides the implications of these results for indirect detection estimates, 
we show that the gravitational effects of such a dark matter spike are significantly smaller than the relativistic effects of the black hole, including frame dragging and quadrupolar effects, for stars orbiting close to the black hole that might be candidates for testing the black hole no-hair theorems. 

\end{abstract}

\maketitle

\section{Introduction and summary}
\label{sec:intro}
The distribution of dark matter in the centers of galaxies is a subject of great interest for several reasons. If, as suggested by N-body 
simulations~\cite{Kuhlen:2012ft}, the density has a cusp at the center because of the large gravitational potential well there, the rates of either decays or annihilations of the dark-matter particles would be enhanced, leading to potentially detectable fluxes of high-energy radiation.  Indeed, our own galactic center has been a key target for ``indirect searches'' for signatures of dark matter~\cite{Bergstrom:1997fj,Bertone05}. The unprecedented accuracy
of the high-energy sky survey provided by the Fermi Large Area 
Telescope (LAT)~\cite{Atwood:2009ez}, and the recent claims for the 
presence of a $130$ GeV line-like 
feature~\cite{Bringmann:2012vr} and of an excess emission
at GeV energies~\cite{Hooper:2011ti} have fuelled a 
sustained interest in the region 
of the galactic center.  In addition, the augmented dark-matter mass distribution could have an influence on the orbits of stars and other matter in the center of the galaxy.

Furthermore, if a massive black hole also resides at the center of the galaxy, its strong gravity could lead to a significant increase in the central region with the creation of a ``spike'' in the dark matter density. 
In their seminal 1999 paper, Gondolo and Silk (\cite{GS}, hereafter referred to as GS) presented a simple model for estimating the density of dark matter in the vicinity of a massive black hole.   Starting with a pre-existing dark-matter density profile, they imagined ``growing'' a massive black hole adiabatically, that is on a timescale long compared to the orbital timescale of a typical dark matter particle.  The phase-space distribution that is implied by the initial density profile evolves in such a way that it retains its form as a function of the relevant dynamical variables, such as energy and angular momentum, but the energy and angular momentum evolve from their initial forms determined by the dynamics of the initial density profile to final forms determined by the dynamics of the dominant black hole in a way that holds the adiabatic invariants of the motion constant.  
Far from the black hole, where gravity is dominated by the dark matter mass distribution, nothing changes.  But close to the black hole the density distribution is significantly modified.

But a black hole is a general relativistic object.  GS carried out an analysis that was primarily Newtonian, but they attempted to take general relativity into account in an {\em ad hoc} way, by adopting a critical angular momentum per unit mass $L_c = 4Gm$ as the minimum value possible for any dark matter particle, where $m$ is the mass of the black hole, $G$ is Newton's constant, and we use units where the speed of light $c=1$.  This is the value such that a marginally bound particle (with relativistic energy per unit mass ${\cal E} = 1$) with this angular momentum will be captured by the black hole.   They found that the density would be strongly increased by the black hole, but that it would vanish generically at a distance $r = 8Gm$, or $4$ Schwarzschild radii from the black hole.  

We have been motivated to re-examine the GS analysis for several reasons.   One of us recently proposed \cite{cwnohair} that, should a number of stars be discovered very close (within a few tenths of a milliparsec) to the black hole Sgr A$^*$ at the center of our galaxy, then future high-precision infrared astronomy capabilities could provide tests of the so-called ``no-hair'' theorem of general relativistic black holes.  The idea would be to measure the precessions of the orbital planes of such stars induced by a combination of the frame dragging and quadrupolar gravity of the rotating black hole, and thereby test the condition required by the Kerr geometry, $Q = - J^2/m$, where $Q$ is the black-hole quadrupole moment and $J$ is its angular momentum.  Whether such as test is feasible depends in part on whether other sources of perturbations of the orbit of a star so close to the black hole would swamp the general relativistic signal.   
Using both semi-analytic techniques and full $N$-body simulations, it was shown that, for a range of possible distributions of stars and stellar-mass black holes in the central region within $4$ milliparsecs (mpc), the orbits of the stars closest to the black hole would still be dominated by relativistic effects \cite{mamw1,lalehcliff}.
A spike of dark matter particles would also perturb stellar orbits because of their gravitational influence, and therefore we were motivated to understand what this influence might be.

Another reason is that the ongoing search for indirect evidence of dark matter at the center of our galaxy~\cite{Bergstrom:1997fj,Bertone05} requires both a model for the density distribution of dark matter at the center as well as models for the various decay~\cite{Bertone:2007aw} or annihilation~\cite{Cirelli:2010zz} processes for various kinds of dark matter.   There are uncertainties in all aspects of these models.   However one thing is certain: if the central black hole Sgr A$^*$ is a rotating Kerr black hole and if general relativity is correct, its external geometry is precisely known.  It therefore makes sense to make use of this certainty as much as possible.

Accordingly, this paper endeavors to put the GS calculation of the dark matter distribution near a black hole on a firm general relativistic footing.   
We begin by incorporating the 
general relativistic effects of a rotating BH exactly in the formal treatment of the number density of particles calculated from a phase space approach.  In order to treat the GS case, we then specialize to a Schwarzschild black hole, and treat the effects of the capture of particles by the hole exactly, show the exact region of integration in energy - angular momentum (${\cal E} - L$) space, and use the exactly formulated adiabatic invariants of the Schwarzschild geometry to show how to grow the black hole using the GS {\em ansatz}.  

\begin{figure}[t]
\begin{center}

\includegraphics[width=4in]{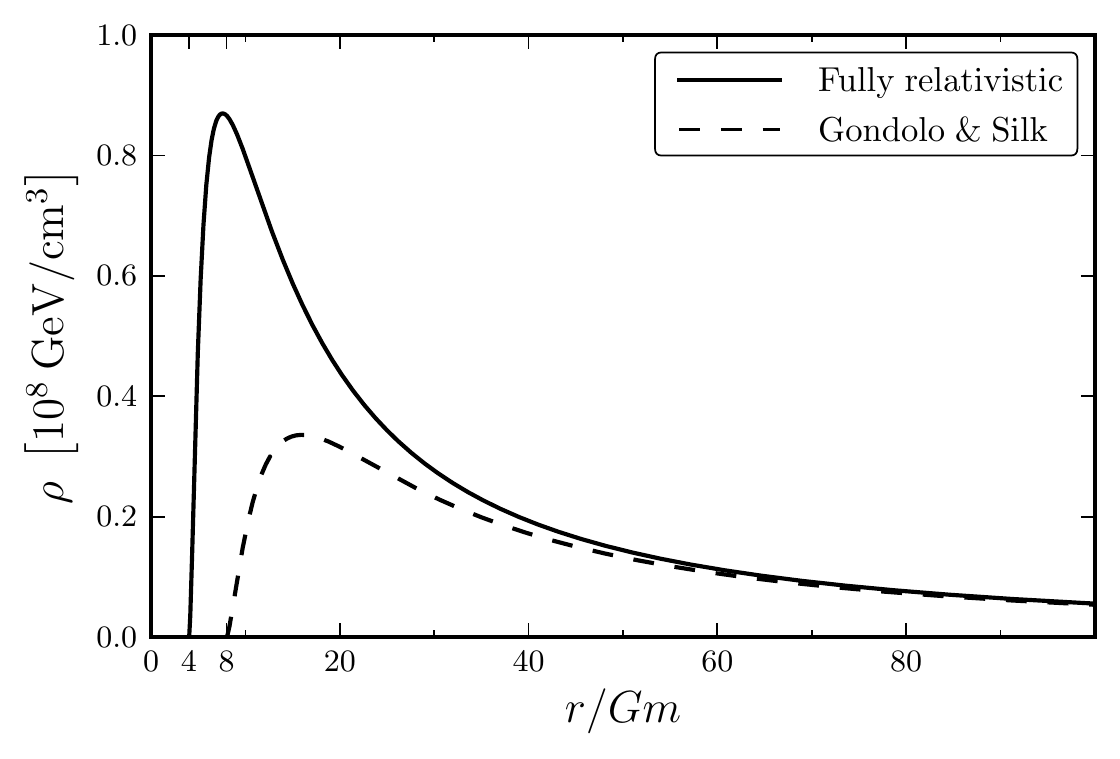}

\caption{\label{fig:plotf0} Number density around a Schwarzschild black hole for a distribution function $f(p)= f_0 =$ constant.  Shown are the fully relativistic and the GS results. }
\end{center}
\end{figure}

Far from the black hole, the resulting density profile matches the profile obtained by GS, and by other workers.  But close to the black hole, there is a significant difference.  This is illustrated in Fig. \ref{fig:plotf0}, which shows the dark matter density $\rho$ for the simple case of a phase space-density $f(p)$ that is constant. This provides an approximation to an initial cored dark matter distribution~\cite{GS,peebles,Young}, $f_0 = \rho_0 \left(2 \pi \sigma_v^2\right)^{-3/2}$, and we
chose typical values for the Milky Way $\rho_0 = 0.3 \: \mathrm{GeV/cm^3}$ and 
$\sigma_v = 100 \: \mathrm{km/s}$. For this case, GS obtained the analytic formula $\rho(r) \propto r^{-3/2} (1 - 8Gm/r)^{3/2}$, implying that the density should vanish at $4$ times the Schwarzschild radius $R_{\rm S} = 2Gm$ .  However we find that the density vanishes at $4Gm$, or twice the Schwarzschild radius, and that the peak density is significantly higher near the black hole than that obtained by GS.

The vanishing of the density at $4Gm$ is simple to understand.  This is the radius of the unstable circular orbit in the Schwarzschild geometry for a marginally bound particle ${\cal E} =1$, with angular momentum per unit mass $L = 4Gm$.  A particle with $L \ge 4Gm$ and ${\cal E} \le 1$ has an inner turning point at $r \ge 4Gm$.  Therefore any particle that manages to reach $r = 4Gm$, i.e. one with precisely ${\cal E} =1$ and $L = 4Gm$ is necessarily captured by the black hole.  Thus the density of particles must vanish at $r=4Gm$, and not at $r=8Gm$.   The attempt by GS to take the capture by the black hole into account did not fully reflect the relativistic nature of the Schwarzschild effective potential.  

Even though the constant phase space density has been used to approximate a dark matter profile with a constant core, the results of N-body 
simulations~\cite{Kuhlen:2012ft} suggest that
the initial dark matter profile is likely to be cusped, behaving as $1/r^\gamma$ close to the center. Hence, we carry out the full adiabatic growth calculation using a Hernquist profile as the initial dark matter density, with a total mass of $10^{12} M_\odot$ and a scale radius of $a = 20$ kpc.  At distances greater than about $50 \, R_{\rm S}$, the density matches standard results; close to the black hole there is a spike $2-3$ times higher than the GS spike, and the density vanishes at $r=4Gm$.  We also take into account the possibility that the dark matter particles are self-annihilating; this leads to a modified profile, with a constant inner density that depends on the annihilation cross-section, the mass of the dark matter particle and the age of the black hole, matching to a Hernquist profile at large distances (see Fig. \ref{fig:hernquist}).   

For both the self-annihilating and non-self-annihilating cases, we calculate the pericenter advance that would be experienced by both the star $S2$, which has recently been seen to complete a full orbit around Sgr A$^*$ (semi-major axis $\sim 4$ mpc), and a hypothetical star in a high-eccentricity orbit near the black hole (semi-major axis $\sim 0.2$ mpc) that might be a candidate for a no-hair-theorem test.
We find that, for both types of dark matter, the pericenter advance induced on a no-hair target star is negligible compared to the relativistic angular precessions expected from the black hole.   For the $S2$ star, the precession is negligible for self-annihilating dark matter, and at the level of a few micro-arcseconds per year as seen from Earth for non-self-annihilating dark matter, such as axions.  This is undetectable at present, but could be marginally detectable with future astrometric capabilities.

The rest of this paper provides the details supporting these conclusions.  In Sec.\ \ref{sec:newton} we review the GS approach using Newtonian theory, and in Sec.\ \ref{sec:phasespace} we carry out the fully general relativistic calculation.   Section \ref{sec:hernquist} applies both the Newtonian and relativistic approaches to the example of an initial Hernquist profile.  In Sec.\ \ref{sec:pericenter} we use the results to estimate the pericenter advance for stars orbiting the galactic center black hole.
Section \ref{sec:conclusion} presents concluding remarks.   Because we work in both Newtonian gravity and general relativity, we use slightly non-standard units, keeping Newton's constant $G$, but setting the speed of light $c=1$.

\section{Growing a black hole in a dark matter cluster: Newtonian analysis}
\label{sec:newton}

We begin with a purely Newtonian analysis of the process of growing a black hole adiabatically within a pre-existing halo, assumed to be isotropic in both position
and velocity space.   
We will generally follow the approach used by Binney and Tremaine~\cite{binneytremaine} (BT hereafter) and Quinlan {\em et al.}~\cite{Quinlan:1994ed}.
In addition to reproducing the non-relativistic results in~\cite{Quinlan:1994ed},
which extended the study of the isothermal sphere carried out in~\cite{Young}, this will set the stage for our fully general relativistic analysis.

Given a distribution function $f(E,L)$, which is normalized to give the
total mass $M$ of the halo upon integration over phase-space, the physical mass density 
is given by:
\be
\rho = \int{  f(E,L)  d^3 \bm{v} } \,,
\ee
where the energy and angular momentum per unit mass $E$ and $L \equiv |{\bm L}|$ are  functions of velocity and position, defined by
\begin{align}
	{\bm L} &= {\bm x} \times {\bm v} \,, \nonumber \\
	E &= \frac{v^2}{2} + \Phi (r)\,, 
\end{align}
where $\Phi (r)$ is the Newtonian gravitational potential.
We now
change integration variables from $\bm{v}$ to $E$, $L$, and the $z$-component of angular-momentum $L_z$, using the relation 
\begin{equation}
d^3 v = J^{-1} dE \,dL\, dL_z \,,
\end{equation}
where the Jacobian is given by the determinant of the matrix
\begin{eqnarray}
J \equiv \left |\frac{\partial (E,\,L,\,L_z)}{\partial (v^x,\, v^y, \, v^z)} \right |
&=&\frac{r}{L} \left|
\begin{array}{ccc}
   v^x&v^y&v^z\\
   (rv^x-x\dot{r})&(rv^y-y\dot{r})&(rv^z-z\dot{r})\\
   -y&x&0\\
\end{array}
\right | \\
\nonumber 
&=&  \frac{r^4}{L} v_r v^\theta \sin \theta \,,
\end{eqnarray}
where $v_r \equiv {\bm x} \cdot {\bm v} /r = \dot{r}$, and 
\begin{equation}
v^\theta = \frac{1}{r^2} {\bm v} \cdot {\bm e}_\theta = \frac{z \dot{r} - r v^z}{r^2 \sin \theta} = \frac{1}{r^2} (L^2 -  L_z^2 \sin^{-2} \theta )^{1/2} \,.
\end{equation}
Including a factor of 4 to take into account the $\pm$ signs of $v^\theta$ and $v_r$ available for each value of $E$ and $L$, we obtain
$d^3 v =4L /(r^4 |v_r| |v^\theta| \sin \theta) dE dL dL_z$, and thus the physical density
\be
\rho (r) = 4 \int {d E} \int
{L d L  \int dL_z \frac{ f(E,L)}{r^4 |v_r| |v^\theta| \sin \theta}} \,.
\label{rhoNewt0}
\ee

We will assume throughout that the distribution function is independent of $L_z$; as a result we can integrate over $L_z$ between the limits $\pm L \sin \theta$, to obtain
Eq.~(1) in~\cite{Quinlan:1994ed}:
\be
\rho (r) = 4 \pi \int {d E} \int
{L d L \frac{ f(E,L)}{r^2 |v_r|}}.
\ee
The limits of integration are set in part by the fact that $|v_r| = (2E-2\Phi - L^2/r^2)^{1/2}$ must be real, and thus $L$ ranges from $0$ to $[2r^2(E-\Phi)]^{1/2}$, while $E$ ranges from $\Phi(r)$ to $0$, the maximum energy that a bound particle could have.  
We thus have 
\be
\rho (r) = \frac{4 \pi}{r^2} \int_{\Phi(r)}^{0} {d E} \int_0^{L_{\rm max}}
{L d L \frac{ f(E,L)}{ \sqrt{2E - 2\Phi(r) - L^2/r^2}}}.
\label{eq:density}
\ee
So given an initial distribution function $f'(E',\,L')$, which acts as a source for the gravitational potential $\Phi(r)$, we can generate the density $\rho(r)$ from
eq.~(\ref{eq:density}). It is often the case that we have knowledge of the
initial dark matter density, $\rho(r)$, either from observations or from fits
to the results of numerical simulations. For a chosen initial $\rho(r)$,
an isotropic distribution function can be constructed by Eddington's method (see BT for discussion).

We next imagine a point mass growing adiabatically at the center of the distribution.   As the gravitational potential near the point mass changes, each particle responds to the change by altering its energy $E$ and angular momentum $L$ and $L_z$, holding the adiabatic invariants $I_r$, $I_\theta$ and $I_\phi$ fixed, where 
\begin{eqnarray}
I_r (E,L) &\equiv& \oint v_r dr = \oint dr \sqrt{2E-2\Phi - L^2/r^2}  \,,
\nonumber \\
I_\theta (L, L_z)&\equiv& \oint v_\theta d\theta = \oint d \theta \sqrt{L^2 - L_z^2 \sin^{-2} \theta }  = 2\pi (L-L_z) \,,
\nonumber \\
I_\phi (L_z) &\equiv& \oint v_\phi d\phi = \oint L_z d\phi = 2\pi L_z \,.
\end{eqnarray}
The constancy of $I_\theta$ and $I_\phi$ implies that $L$ and $L_z$ remain constant, no surprise considering the assumed spherical symmetry.   But when the potential evolves from the initial potential $\Phi'$ to a new potential $\Phi$ that includes the point mass, $E'$ evolves to $E$ such that
\begin{equation}
I_r (E,\, L) = {I'}_r (E',\, L) \,.
\label{radialactionequal}
\end{equation}
As shown in~\cite{Young} (and extended to the general relativistic domain in~\cite{lalehthesis}), the distribution function is invariant under adiabatic evolution, $f(E,L) = f'(E'(E,L), L)$; in the original distribution function, $E'$ is expressed in terms of $E$ and $L$ by inverting Eq.~(\ref{radialactionequal}).   Note that, for a potential dominated by a point Newtonian mass, $\Phi = -Gm/r$, and
\begin{equation}
I_r (E,\, L) = 2\pi \left ( -L + \frac{Gm}{\sqrt{-2E}} \right ) \,.
\label{eq:pointradialaction}
\end{equation}
The density in the presence of the growing point mass may then be expressed as 
\be
\rho (r) = \frac{4 \pi}{r^2} \int_{-Gm/r}^{0} {d E} \int_0^{L_{\rm max}}
{L d L \frac{ f'(E'(E,L),L)}{\sqrt{2E + 2Gm/r - L^2/r^2}}}.
\ee

\section{Growing a black hole in a dark matter cluster: Relativistic analysis} 
\label{sec:phasespace} 

%\subsection{General considerations}

Given a system of particles characterized by a distribution function $f^{(4)}(p)$,  there is a standard prescription for writing down the mass current four-vector~\cite{fackerell,shapiroteukolsky}:
\begin{equation}
J^{\mu} (x) \equiv \int f^{(4)}(p) \frac{p^\mu}{\mu} \sqrt{-g} \,d^4p \,,
\end{equation}
where $\mu$ is the particle's rest mass,  $p$ and $p^\mu$ represent the four-momentum, $g$ is the determinant of the metric, and $d^4p$ is the four-momentum volume element; the distribution function is again normalized so that the total mass of the halo is $M$.  

As in the Newtonian case, we wish to change variables from $p^\mu$ to variables that are related to suitable constants of the motion.  In the absence of a black hole, and for a spherically symmetric cluster, the constants would be the relativistic energy $\cal E$, the angular momentum and its $z$-component $(L, \, L_z)$, together with the conserved rest-mass $\mu = (-p_\mu p^\mu)^{1/2}$.    A black hole that forms at the center will generically be a Kerr black hole, whose constants of motion are ${\cal E}$, $L_z$, $\mu$, plus the so-called Carter constant $C$.   In the limit of spherical symmetry, such as for no black hole or for a central Schwarzschild black hole, $C \to L^2$.    

We will therefore begin by changing coordinates in the phase-space integral from $p^\mu$ to ${\cal E}$, $C$, $L_z$ and $\mu$ assuming that the background geometry is the Kerr spacetime.   We will find that the loss of spherical symmetry and the dragging of inertial frames that go together with the Kerr geometry make the problem considerably more complex.  Further study of this case will be deferred to future work.  Taking the limit of a Schwarzshild black hole simplifies the analysis, and allows us to formulate the adiabatic growth of a non-rotating black hole in a fully relativistic manner. 

\subsection{Kerr black hole background}

The Kerr metric is given in Boyer-Lindquist coordinates by
\begin{eqnarray}
ds^2 &=& - \biggl ( 1 - \frac{2Gmr}{\Sigma^2} \biggr ) dt^2
 + \frac{\Sigma^2}{\Delta} dr^2 + \Sigma^2 d\theta^2
 - \frac{4Gmra}{\Sigma^2} \sin^2 \theta dt d\phi
 \nonumber \\
 && \quad + \biggl ( r^2 + a^2 + \frac{2Gmra^2 \sin^2 \theta}{\Sigma^2} \biggr ) \sin^2 \theta d\phi^2  \,,
\end{eqnarray}
where $G$ is Newton's constant, $m$ is the mass, $a$ is the Kerr parameter, related to the angular momentum $J$ by 
$a \equiv J/m$; $\Sigma^2 = r^2 + a^2 \cos^2 \theta$, and 
$\Delta = r^2 + a^2 - 2Gmr$.  We will assume throughout that $a$ is positive, and use units in which $c=1$.

Timelike geodesics in this geometry admit four conserved quantities: energy  of the particle per unit mass, ${\cal E}$; angular momentum per unit mass,  $L_z$; Carter constant per unit (mass)$^2$,  $C$; and the norm of the four momentum: 
\begin{subequations}
\begin{eqnarray}
{\cal E} &\equiv& - u_0 = -g_{00} u^0 - g_{0\phi} u^\phi \,,
\label{energy}
\\
L_z &\equiv&  u_\phi = g_{0\phi} u^0 + g_{\phi \phi} u^\phi \,,
\label{Lz}
\\
C &\equiv& \Sigma^4 \bigl ( u^\theta \bigr )^2 + \sin^{-2} \theta L_z^2 + 
a^2 \cos^2 \theta ( 1 - {\cal E}^2 ) \,,
\label{carter}
\\
g_{\mu\nu} p^\mu p^\nu &=& -\mu^2 \,.
\label{norm}
\end{eqnarray}
\label{constants}
\end{subequations}
The version of the Carter constant used here has the property that, in the Schwarzschild limit ($a \to 0$), $C \to L^2$, where $L$ is the total conserved angular momentum per unit mass.

We want to convert from the phase space volume element $d^4p$ to the volume element $d{\cal E} dC dL_z d\mu$, using the relation
\begin{equation}
d^4 p = |J|^{-1} d{\cal E} dC dL_z d\mu \,,
\end{equation}
where the Jacobian is given by the determinant of the matrix
\begin{eqnarray}
J \equiv \left |\frac{\partial ({\cal E},\,C,\,L_z,\,\mu)}{\partial (p^0,p^r,p^\theta,p^\phi)} \right |
&=&\mu^{-3} \left|
\begin{array}{cccc}
   -g_{00}&0&0&-g_{0\phi}\\
   \partial C/\partial u^0&0&2 \Sigma^4 u^\theta&\partial C/\partial u^\phi \\
   g_{0\phi}&0&0&g_{\phi\phi}\\
   {\cal E}&-u_r&-u_\theta&-L_z\\
\end{array}
\right | \\
\nonumber 
&=&  -2 \mu^{-3} \Delta \Sigma^4 u_r u^\theta \sin^2 \theta \,.
\end{eqnarray}
Again including a factor of 4 to take into account the $\pm$ signs of $p^\theta$ and $p^r$ in contrast to the quadratic nature of $C$ and the norm of $p^\mu$, and using the fact that $\sqrt{-g} = \Sigma^2 \sin \theta$, we obtain
\begin{equation}
\sqrt{-g} \, d^4p = \frac{2 \mu^3 }{\Sigma^2 \Delta |u_r| |u^\theta| \sin \theta} d{\cal E} dC dL_z d\mu  \,.
\end{equation}
If the particles in the distribution have the same rest mass, and if we again assume that the three-dimensional distribution function is normalized as before, then $f^{(4)}(p) \equiv \mu^{-3} f({\cal E}, C) \delta (\mu - \mu_0)$, and thus we can integrate over $\mu$, to obtain 
\begin{equation}
J^\mu = 2 \int d {\cal E} \int
d C  \int dL_z   \frac{u^\mu f({\cal E},C) }{\Sigma^2 \Delta  |u_r| |u^\theta| \sin \theta} \,.
\label{masscurrent}
\end{equation}
We again assume that $f$ is independent of $L_z$.
Equation (\ref{masscurrent}) may be compared with Eq.~(\ref{rhoNewt0}); $J^0$ is related to the density $\rho$, the relativistic energy $\cal E$ replaces $E$, $C$ plays the role of $L^2$, $\Sigma^2 \Delta$ replaces $r^4$, and four-velocities $u_r$ and $u^\theta$ replace ordinary velocities $v_r$ and $v^\theta$.  

By definition, $J^\mu \equiv \rho u^\mu$, where $\rho$ is the mass density as measured in a local freely falling frame, and $u^\mu$ is the four-velocity of an element of the matter, which can be expressed in the form $u^\mu \equiv \gamma (1, v^j)$, where $v^j \equiv u^j/u^0 = J^j /J^0$, and $\gamma \equiv (-g_{00} - 2 g_{0j} v^j - g_{ij} v^i v^j )^{-1/2}$.  Thus, once the components of $J^\mu$ are known, then $u^0 = \gamma$ can be determined, and from that $\rho = J^0/u^0$ can be found.  Alternatively, because the norm of $u^\mu$ is $-1$, $\rho = (-J_\mu J^\mu)^{1/2}$.   In particular, if $J^\mu$ has no spatial components, then $u^0 = (-g_{00})^{-1/2}$ and $\rho = \sqrt{-g_{00}} J^0 $.

The four-velocity components $u_r$ and $u^\theta$ can be expressed in terms of the constants of the motion by suitably manipulating Eqs.\ (\ref{carter}) and (\ref{norm}), leading to
\begin{eqnarray}
u^\theta &=& \pm \Sigma^{-2} \left [ C -  L_z^2 \sin^{-2} \theta - a^2 \cos^2 \theta (1 - {\cal E}^2)  \right ]^{1/2} \,,
\nonumber \\
u_r &=& \pm \frac{r^2}{\Delta} V(r)^{1/2} \,,
\end{eqnarray}
where
\begin{equation}
V(r) =  \biggl ( 1+ \frac{a^2}{r^2} +\frac{2Gma^2}{r^3}  \biggr ){\cal E}^2 -\frac{\Delta}{r^2} \biggl (1 + \frac{C}{r^2} \biggr ) + \frac{a^2 L_z^2}{r^4} - \frac{4Gma{\cal E}L_z}{r^3} \,.
\label{Vr}
\end{equation}

From Eq.~(\ref{masscurrent}), it is clear that, since $u^r$ and $u^\theta$ are equally likely to be positive as negative for a given set of values for $\cal E$, $C$ and $L_z$, the components $J^r$ and $J^\theta$ of the current must vanish.
Furthermore, since $u_0 = -{\cal E}$ and $u_\phi = L_z$, we have that 
\begin{eqnarray}
J_0 &=& -2 \int {\cal E} d {\cal E} \int
d C  \int dL_z   \frac{f({\cal E},C) }{\Sigma^2 \Delta  |u_r| |u^\theta| \sin \theta} \,,
\nonumber  \\
J_\phi &=& 2 \int  d{\cal E} \int
d C  \int  L_z dL_z   \frac{f({\cal E},C) }{\Sigma^2 \Delta  |u_r| |u^\theta| \sin \theta} \,.
\label{eq:J0}
\end{eqnarray}

Even if we assume that $f$ is independent of $L_z$, the presence of the term in $V(r)$ [Eq.~(\ref{Vr})] that is {\em linear} in $L_z$ implies that $J_\phi$ will {\em not} vanish in general, and thus the distribution of matter will have a flux in the azimuthal direction.  This, of course, is the dragging of inertial frames induced by the rotation of the black hole, an effect that will be proportional to the Kerr parameter $a$.   In this case the density may be obtained from
\begin{eqnarray}
\rho &=& (- g^{00} J_0^2 - 2 g^{0\phi} J_0 J_\phi - g^{\phi \phi} J_\phi^2 )^{1/2} 
\nonumber \\
&=& - J_0 \left ( \frac{g_{\phi \phi} + 2 g_{0\phi} \Omega + g_{00} \Omega^2 }{\Delta} \right )^{1/2} \,,
\end{eqnarray}
where $\Omega \equiv J_\phi/J_0$.    If $a=0$, then $J_\phi = 0$, and $\rho = -J_0 (g_{\phi\phi}/\Delta)^{1/2}
= -J_0 (-g^{00})^{1/2} = \sqrt{-g_{00}} J^0$.

The three-dimensional region of integration over $\cal E$, $C$ and $L_z$ is complicated.  The energy ${\cal E}$ is bounded above by unity if unbound particles are to be excluded from consideration.  The variables are bounded by the two-dimensional surfaces defined by $u^\theta =0$ and $u_r =0$, the latter depending on the value of $r$.  A final bound is provided by the condition that if a given particle has an orbit taking it close enough to the black hole to be captured, it will disappear from the distribution.  For a given $\cal E$ and $L_z$ there is a critical value of $C$, below which a particle will be captured.   No analytic form for this condition has been found to date, although for non-relativistic particles for which ${\cal E} =1$ is a good approximation, Will~\cite{willcapture} found an approximate analytic expression for the critical value of $C$.

The adiabatic invariants in this case are given by
\begin{eqnarray}
I_r ({\cal E}, C, L_z) &\equiv& \oint u_r dr = \oint dr V(r)^{1/2} \left (1-\frac{2Gm}{r} + \frac{a^2}{r^2} \right)^{-1}   \,,
\nonumber \\
I_\theta ({\cal E}, C, L_z)&\equiv& \oint u_\theta d\theta = \oint d \theta \left [ C -  L_z^2 \sin^{-2} \theta - a^2 \cos^2 \theta (1 - {\cal E}^2)  \right ]^{1/2} \,,
\nonumber \\
I_\phi (L_z) &\equiv& \oint u_\phi d\phi =  2\pi L_z \,.
\label{adiabat_rel}
\end{eqnarray}

\subsection{Schwarzschild black hole background}

We now restrict our attention to the Schwarzschild limit, $a = 0$, in which $\Sigma^2 = r^2$, $C= L^2$, $u^\theta = (L^2 - L_z^2 \sin^{-2} \theta)^{1/2}$ and 
\begin{equation}
V(r) = {\cal E}^2 - \biggl ( 1 - \frac{2Gm}{r} \biggr ) \biggl ( 1 + \frac{L^2}{r^2} \biggr ) \,.
\end{equation}
The metric components are $g_{00} = -g_{rr}^{-1} = -1 + 2Gm/r$, and $g_{0\phi} =0$.  Substituting these relations into  Eqs.~(\ref{eq:J0}), we write $J_0$ in the form
\begin{equation}
J_0 = -\frac{2}{r^2} \int {\cal E} d {\cal E} \int
d L^2  \int dL_z   \frac{f({\cal E},L) }{V(r)^{1/2} (L^2 \sin^2 \theta - L_z^2)^{1/2} } \,,
\end{equation}
and we observe that $J_\phi = 0$. 
We then integrate over $L_z$ explicitly  to obtain
\begin{equation}
J_0 = -\frac{4\pi }{r^2} \int {\cal E} d {\cal E} \int L
d L \frac{f({\cal E},L) }{\sqrt{{\cal E}^2 - (1-2Gm/r)(1 + L^2/r^2)}}  \,. 
\label{J0finalS}
\end{equation}
We again assume that $\cal E$ is bounded from above by unity;  ${\cal E}$ and $L$ are also bounded by the vanishing of $V(r)$ and by the black hole capture condition.

Unlike the Kerr case, the capture condition in Schwarzschild can be derived analytically.  We wish to find the critical value of $L$ such that an
orbit of a given energy ${\cal E}$, and  $L$ will not be ``reflected'' back to large distances, but instead will continue immediately to smaller values of $r$ and be captured by the black hole.  The turning points of the orbit are given by the values of $r$ where  $V(r)=0$.  The critical values of  ${\cal E}$ and $L$ are those for which the potential has an extremum at that same point, that is where $dV(r)/dr = 0$.  The chosen sign for $V(r)$ also dictates that this point should be a minimum of $V(r)$, that is that $d^2 V(r)/dr^2 > 0$, corresponding to an unstable extremum.  
We obtain from the condition $dV(r)/dr =0$ the standard solution for the radius of the unstable circular orbit in Schwarzschild $r = 6Gm/ \{1 + [1- 12(Gm/L)^2]^{1/2}\}$.  Substituting this into the condition $V(r) =0$ and solving for $L$, we obtain the critical value
\begin{equation}
{L}_c^2 = \frac{32(Gm)^2}{36{\cal E}^2 - 27{\cal E}^4 -8 + {\cal E}(9{\cal E}^2-8)^{3/2}} \,.
\label{Lcrit}
\end{equation}
Notice that, for ${\cal E} =1$, $L_c = 4Gm$, corresponding to the unstable marginally bound orbit in Schwarzschild at $r=4Gm$, while for ${\cal E} = (8/9)^{1/2}$, $L_c = 2\sqrt{3}Gm$, corresponding to the innermost stable circular orbit at $r=6Gm$.  

\begin{figure}[t]
\begin{center}

\includegraphics[width=4in]{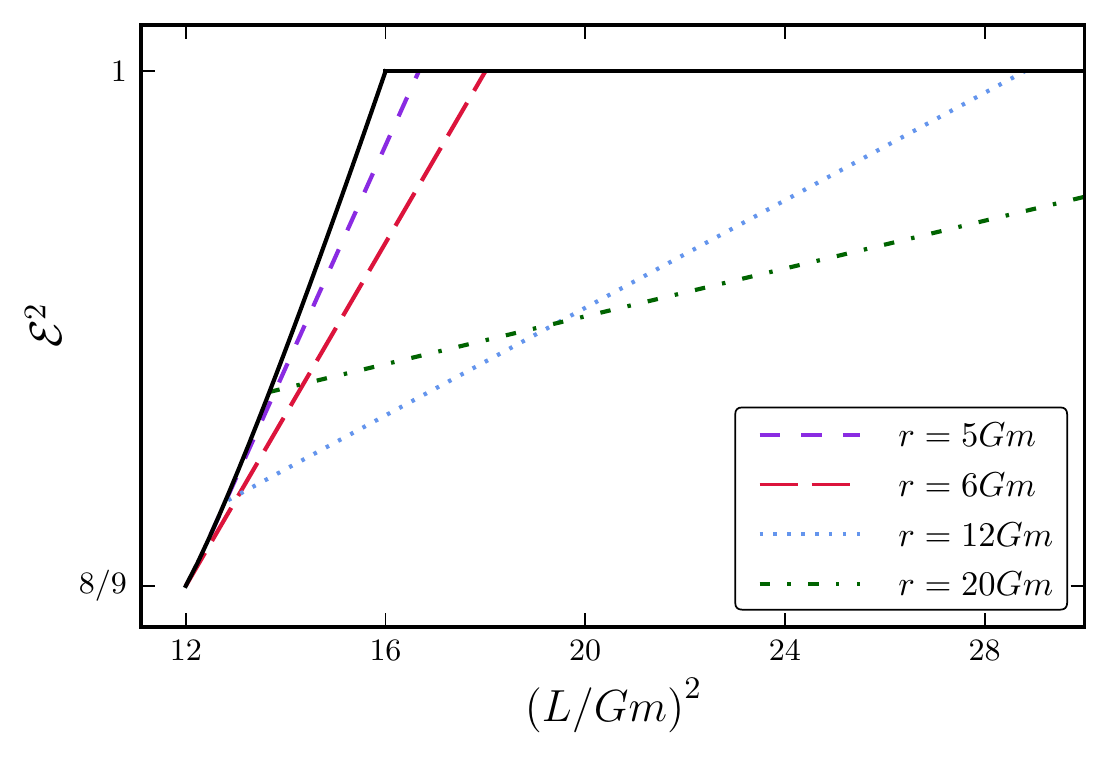}

\caption{\label{fig:phasespace} Integrating over ${\cal E}$-$L$ space for the Schwarzschild geometry.  For a given $r$, the region of integration lies between the solid lines and the various dashed and dotted lines. As $r \to 4Gm$, the integration area vanishes. }
\end{center}
\end{figure}

The range of integration of the variables is therefore as follows:   $L$ is integrated from  $L_{\rm min} = L_c$ to the value given by $V(r) = 0$, namely
\begin{equation}
L_{\rm max} = r \left ( \frac{{\cal E}^2}{1-2Gm/r} -1 \right )^{1/2} \,.
\end{equation}
The energy ${\cal E}$ is then integrated between its minimum value and unity.  That minimum value is found by solving $V(r) =0$ with $L = L_c$, and is given by
\begin{eqnarray}
{\cal E}_{\rm min} &=& \left \{ \begin{array}{ll}
   (1+ 2Gm/r)/(1+6Gm/r)^{1/2}  & : r \ge 6 Gm \\
   (1- 2Gm/r)/(1-3Gm/r)^{1/2} &: 4Gm \le r \le 6Gm \,.\\
    \end{array}
    \right .
    \label{Elimits}
\end{eqnarray}
The regions of integration for various values of $r$ are shown in Fig.\ \ref{fig:phasespace}.  For a given $r$, the region is a triangle bounded by the critical capture angular momentum on the left, the maximum energy ${\cal E} =1$ at the top, and the condition $V(r)=0$ on the triangle's lower edge.   For $r=6Gm$, the lower edge of the region is the long dashed line shown (red in color version).  As $r$ increases above $6Gm$ the lower edge of the triangle moves upward and the right-hand vertex moves rightward, as shown by the dotted and dot-dashed lines in Fig. \ref{fig:phasespace} (blue and green in color version).  For values of $r$ decreasing below $6Gm$, the lower edge of the triangle moves upward and leftward as shown by the short dashed line in  Fig. \ref{fig:phasespace} (violet in color version).   At $r= 4Gm$, ${\cal E}_{\rm min} =  {\cal E}_{\rm max}  =1$ and 
$L_{\rm min} =  L_{\rm max}  =4$, and the volume of phase space vanishes.  This implies that, irrespective of the nature of the distribution function, the density of particles must vanish at $r=4Gm$; this makes physical sense, since any bound particle that is capable of reaching $r=4Gm$ is necessarily captured by the black hole and leaves the distribution.   This is a rather different conclusion from the one reached by GS, who argued that the density would generically vanish at $r=8Gm$.  
The specific shape of this phase space region for small $r$ will play a central role in determining the density distribution near the black hole.

In the Schwarzschild limit, the four-velocity components are given by $u_\phi = L_z$, $u_\theta = (L^2 - L_z^2 \sin^{-2} \theta )^{1/2}$, and $u_r = [{\cal E}^2 - (1-2Gm/r)(1+L^2/r^2)]^{1/2}$, so that the adiabatic invariants are
\begin{eqnarray} \label{Ir_rel}
I_r (E,L) &=& \oint dr \sqrt{{\cal E}^2 - (1-2Gm/r)(1+L^2/r^2)}  \,,
\nonumber \\
I_\theta (L, L_z)&=&  2\pi (L-L_z) \,,
\nonumber \\
I_\phi (L_z) &=& 2\pi L_z \,.
\end{eqnarray}  

\subsection{Example: Constant  distribution function}

To illustrate the application of these results, we consider the special, albeit unrealistic case of a constant distribution function $f({\cal E},L) = f_0$.  Then $f$ is still constant after applying the adiabatic condition.  
Since $f$ is independent of $L$, we can do the $L$ integration explicitly to obtain
\begin{equation}
J_0 = -4\pi  f_0  \int \frac{\sqrt{{\cal E}^2 - (1-2Gm/r)(1 + L_c^2/r^2)}}{1-2Gm/r} {\cal E} d {\cal E}  \,, 
\end{equation}
from which we obtain the density
\begin{equation}
\rho(r) = \frac{4\pi  f_0}{(1-2Gm/r)^{3/2}} \int \sqrt{{\cal E}^2 - (1-2Gm/r)(1 + L_c^2/r^2)} \,
{\cal E} d {\cal E}  \,.
\label{nfinal}
\end{equation}
Substituting Eq.\ (\ref{Lcrit}) for $L_c$ and integrating over ${\cal E}$ numerically between the limits shown in Eq.\ (\ref{Elimits}), we obtain the number density plotted in Fig.\ \ref{fig:plotf0}.  

Gondolo and Silk~\cite{GS} attempted to incorporate the relativistic effects of the black hole within a Newtonian context as follows.  First they approximated the energy $\cal E$ by ${\cal E} = 1 + E$ with $E <0$, so that, to Newtonian order,
the denominator in Eq.~(\ref{J0finalS}) is $\approx [2(E + Gm/r) - L^2/r^2]^{1/2}$, and ${\cal E} d{\cal E} \approx dE$.  For the critical capture angular momentum they adopted the approximation $L_c = 4Gm$, the value corresponding to ${\cal E} =1$, while for the minimum energy, they adopted the value of $E$ for which the denominator vanishes for {\em that} critical angular momentum.  For the constant distribution function the integrals can be done analytically, with the result [GS, Eq.\ (6)]
\begin{equation}
\rho(r) = \frac{4\pi f_0}{3} \left ( \frac{2Gm}{r} \right )^{3/2} \left ( 1 - \frac{8Gm}{r} \right )^{3/2} \,.
\label{rhoGS}
\end{equation}
In Fig.\ \ref{fig:plotf0} we plot Eq.\ (\ref{rhoGS}) for comparison with the relativistic result.  The two distributions agree completely at large distances, as expected.   The GS distribution vanishes at $r = 8Gm$, and is a factor of three smaller at its peak than the fully relativistic distribution.   Interestingly, the simple replacement of $8Gm$ by $4Gm$ in the GS formula gives a distribution with the correct behavior at short and large distances, and peaks at a value about 15\% higher than our numerical result.

\section{Application: the Hernquist model}
\label{sec:hernquist}

\subsection{Newtonian analysis}

As an example of models with an inner cusp, favored by the results of N-body
simulations, we consider a Hernquist profile~\cite{Hernquist:1990be}.
The Hernquist model is a spherically symmetric matter distribution whose density and  Newtonian gravitational potential are given by
\begin{equation}
 \rho(r) = \frac{\rho_0}{(r/a)(1+r/a)^3} \,, \quad \Phi = - \frac{GM}{a+r} \,,
 \label{hernquist0}
\end{equation}
where $\rho_0$ and $a$ are scale factors, and
$M \equiv 2\pi \rho_0 a^3$ is the total mass of the cluster. 
Although the Hernquist profile describes isolated dark matter 
halos~\cite{Visbal:2012ta}, because of the continued infall
from the cosmological background, halos found in cosmological simulations
are better fit by an NFW profile~\cite{Navarro:1996gj}. 
Our choice of the Hernquist profile is motivated by the fact that it captures the
same behavior at distances close to the center as the NFW profile, $\propto 1/r$,
but it has the advantage that its associated ergodic distribution function can be 
found analytically. As shown in~\cite{Quinlan:1994ed}, both profiles fall in
the class of $\gamma$-models resulting in the same Newtonian spike.

The distribution function that is consistent with this potential is given by the  (properly normalized)
Hernquist form 
\be
f_H \left( \te \right) = \frac{M}{\sqrt{2} (2 \pi)^3 (G M a)^{3/2}} 
\tilde{f}_H \left(\te \right) \,,
\label{hernquist1}
\ee
where 
\be
\tilde{f}_H \left(\te \right) = \frac{\sqrt{\te}}{\left(1-\te\right)^2}
\left[ \left(1 - 2 \te \right) \left(8 \te^2 - 8 \te - 3 \right)+
\frac{3 \sin^{-1} \sqrt{\te}}{\sqrt{\te \left(1-\te\right)}} \right] \,,
\label{hernquist2}
\ee
where we adopt the following 
dimensionless quantities:
\begin{align}
	{\te}& \equiv - \frac{a}{G M}E \,, \notag \\
	\tilde{L} & \equiv \frac{L}{\sqrt{a G M}} \,,\notag \\
	x & \equiv r/a \,,\notag \\
	\tp &\equiv - \frac{a}{G M} \Phi = \frac{1}{1+x} \,, \notag \\
	\tilde{m} & \equiv m /M \,,
	\label{eq:dimensions}
\end{align}
where $m$ is the mass of the black hole.

With these definitions, Eq.~(\ref{eq:density}) for the density becomes:
\begin{align}
	\rho(r) &= 4 \pi \left( \frac{GM}{a}\right)^{3/2} 
	\int_0^{\te_{\rm max} (x)} { d \te} \int_{\tl_{\rm min}}^{\tl_{\rm max}}{ \tl d \tl
		\frac{f_H(\te)}{x^2 \sqrt{2 \left(\tp -\te 
	\right) - \tls/x^2}}} \nonumber \\
	& = \frac{1}{ \sqrt{2} (2\pi)^2 x} \left (\frac{M}{a^3} \right )
	\int_0^{\te_{\rm max} (x)} { d \te} \int_{\tl_{\rm min}^2}^{\tl_{\rm max}^2}{ d \tl^2
		\frac{\tilde{f}_H(\te)}{\sqrt{\tls_{\rm max}- \tls}}} \,,
	\label{eq:densityx}
\end{align}
where $\tl_{\rm max}^2 = 2x^2(\tp - \te)$ and $\tilde{f}_H(\te)$ is given by Eq. (\ref{hernquist2}).   Normally we would have $\tl_{\rm min} =0$,  and 
$\te_{\rm max} (x) = \tp (x)$.  But we will allow the more general limits in order to include for comparison the GS ansatz for incorporating black-hole capture effects, namely $\tl_{\rm min} = 4\tm (GM/a)^{1/2}$ and $\te_{\rm max} (x) = \tp (x) (1-8\tm M/xa)$.   

When we now grow a point mass adiabatially within the Hernquist model,
the argument $\te'$ 
of the initial distribution (\ref{hernquist2}) becomes a function of $\te$ and $L$ by 
equating the radial actions:
\be
I_r^H \left(\te{}', \tl \right) = I_r^{\rm bh} \left( \te, \tl \right),
\label{eq:adiabatic}
\ee
and using the fact that $\tl'=\tl$ from the angular action.  
Hence the density around the point mass in a Hernquist profile takes the form:
\be
	\rho(r) = \frac{1}{ \sqrt{2} (2\pi)^2 x} \left (\frac{M}{a^3} \right )
	\int_0^{\tm/x} { d \te} \int_{\tl_{min}}^{\tl_{max}}{ d \tls
		\frac{\tilde{f}_H \left(\te{}'(\te,\tl)\right)}
			{\sqrt{\tls_{\rm max} - \tls}}} \,,
	\label{eq:densityh}
\ee
where $\tls_{\rm max} = 2x^2 (\tm/x - \te)$.

From Eq.~(\ref{eq:pointradialaction}), the radial adiabatic invariant for a point mass potential in dimensionless variables is
\be
I_r^{\rm bh}= 2\pi \sqrt{GMa} \left( \frac{\tm}{\sqrt{2 \te}} - \tl \right) \,.
\ee
We see that it diverges for $\epsilon \rightarrow 0$, corresponding to 
the least bound particle. We will have to be careful when matching the 
radial actions in this limit.

For the Hernquist potential, with 
$\tp = 1/(1+x)$ an analytic formula cannot be found for the radial invariant
\be
I_r^H = 2\sqrt{GMa}  \int_{x_-}^{x_+} \left ( \frac{2}{1+x} - 2\te - \frac{\tls}
{x^2} \right )^{1/2} \, d x \,,
\ee
and thus it will have to be evaluated numerically.  
To this end, it is convenient to transform the integration in
the following way.
First, combine the three terms inside the square root to get
\be
\frac{2}{1+x} - 2\te - \frac{\tls}{x^2} =
\frac{-2 \te x^3 + 2(1- \te) x^2 - \tls x -\tls}{x^2 (1+x)}.
\ee
We solve for the three roots of the numerator, of which the two positive roots give the
turning points $x_+$ and $x_-$, while the third root $x_{\rm neg}$
is always negative.
%and it is the only one that exists whenever we
%are outside of the domain describing bound particles:
%\be
%0 < \Delta \equiv 4 \tls \left(8 + \tls -2 \epsilon \left(\tls + 4 \epsilon
%		\left(\epsilon -3 -\tls\right) + \tls \left(10 + \tls\right)
%\right) \right).
%\ee
We then rewrite the function in
the square root as:
\be
2 \epsilon \frac{(x_+-x)(x-x_-)(x-x_{neg})}{x^2 (x+1)},
\ee
which is positive in the region $x_- \le x \le x_+$.
We now make a change of variables $x = t \left(x_+ - x_-\right) + x_-$,
which brings the integral into the domain $[0,1]$:
\be \label{Ir_H}
I_r^H =2\sqrt{GMa} \sqrt{2 \te} \left(x_+ - x_-\right)^2
\int_0^1 {\sqrt{\frac{(1-t) t \left( (x_+ - x_-) t + x_- -x_{neg} \right)}
{(x_+ - x_-) t + x_-}} \frac{d t}{(x_+ - x_-) t + x_-+1}}.
\ee
This makes it much easier to control the integration numerically, since we can make
sure that the roots have the right signs and ordering, and no numerical
round-off errors will prevent the evaluation of the real square root.

For $\tls=0$, the radial invariant can be integrated analytically, with the turning
points $x_-=0$ and $x_+=1/\epsilon -1$, 
\begin{eqnarray} \nonumber
I_r^H&=&2\sqrt{GMa}\int_0^{1/\te-1} \sqrt{\frac{2}{1+x}-2\te} \ dx \ , \\
&=&2\sqrt{2GMa} \left[ \frac{\arccos{\sqrt{\te}}}{\sqrt{\te}}-\sqrt {1-\te} \right] \ ,
\end{eqnarray}
and we use this fact in the code.
The radial invariant  is again divergent for $\epsilon \rightarrow 0$. Since we are only interested in finding a solution
in the domain $(0,1]$, we simply define the value there to be a very large
number, and use a bracketing algorithm.

For numerical work, it is also convenient to remap the integral (\ref{eq:densityh}) for $\rho(r)$ into a square
domain.  
This is a particular case of a set of transformations discovered by
Duffy~\cite{Duffy}. 
We make a change of variables, $( \te, \tls) \rightarrow
(u,z)$, that maps the domain of integration
in Eq.~(\ref{eq:densityx}) onto the square $[0,1]\times [0,1]$:
\begin{align}
	\te \equiv & u \te_{max} \nn
	\tls \equiv & z \tls_{max}(u) + (1-z) \tls_{min} \,,
	\label{eq:uz}
\end{align}
where we emphasize that $\tls_{max}$ depends on $u$.

The jacobian is:
\begin{align}
	\frac{\left(\partial \te, \partial \tls\right)}{(\partial u, \partial
	z)} &=\left| \begin{array}{cc} \te_{max}  & 0 \\ \ldots & 
		\tls_{max}(u) -
	\tls_{min} \end{array} \right| \nn
	&=\te_{max} \left(\tls_{max}(u) - \tls_{min} \right).
\end{align}

With this change, the integral in Eq.~(\ref{eq:densityx}) reads:
\be
\rho (r) = \frac{1}{ \sqrt{2} (2\pi)^2 x} \left (\frac{M}{a^3} \right ) \te_{\rm max} \int_0^1{d u \int_0^1{d z 
		\sqrt{\frac{\tls_{max}(u) - \tls_{min}}{1-z}}
\tilde{f}_H(\tilde{\epsilon}'(u,z))}},
\label{eq:rhobox}
\ee
where the arguments of the distribution function are given 
in Eq.~(\ref{eq:uz}).   This has the effect of  making our codes faster and more stable. One
of the advantages is that the integrable singularity that was originally
in a corner ($\te = \tp$, $ \tls=0$) of the integration domain has now been
transferred to a line, depending only on the variable $z$. 

Using the GS conditions for $\tilde{L}_{\rm min}$ and $\tilde{\epsilon}_{\rm max}$ and carrying out the numerical integrations, we obtain the curve labeled ``Non-relativistic'' in Fig.\ \ref{fig:hernquist}.

\subsection{Relativistic analysis}

We now apply these considerations to the relativistic formalism. 
Here we define $\te$ in terms of the relativistic energy $\cal E$ per unit particle mass using
\be
\te \equiv \frac{a}{GM}(1-{\cal E}) \,;
\ee
the other definitions in Eqs.~(\ref{eq:dimensions}) will be the same. Using these definitions, and the relation $\rho = -J_0(-g^{00})^{1/2}$ along with Eq.~(\ref{J0finalS}), we find 
\begin{eqnarray} \nonumber
\rho(r)
&=&\frac{4\pi}{x^2}\frac{(GM/a)^{3/2}}{\sqrt{1-2Gm/r}}  
\int_0^{\te_{\rm max}} [1-(GM/a)\te] \ d \te \times 
\\ \nonumber
&& \qquad
\int_{\tl_{\rm min}}^{\tl_{\rm max}} \tl d \tl \frac{f_H(\te)}{\sqrt{2(\tm/x-\te)-\tl^2/x^2+(GM/a)\te^2+(2GM/a)(\tm/x)(\tl^2/x^2)}} \ , \nonumber \\ 
\label{ch3-48}
&=&\frac{1}{ \sqrt{2} (2\pi)^2} \left (\frac{M}{a^3} \right )\frac{a}{r-2Gm} \int_0^{\te_{\rm max}} [1-(GM/a)\te] \ d \te \int_{\tls_{\rm min}}^{\tls_{\rm max}} d \tls \  \frac{\tilde{f}_H(\te)}{\sqrt{\tls_{\rm max}-\tls}} \ ,
\end{eqnarray}
where $\tilde{f}(\te)$ is again given by Eq. (\ref{hernquist2}), and 
where we used ${\cal E}=1$ for the maximum energy of the bound particles which leads to $\te_{\rm min}=0$.  Compare the last equation of (\ref{ch3-48}) to Eq. (\ref{eq:densityx}).  

To consider the growth of the central black hole and its capture effects, we use Eqs. (\ref{Lcrit})-(\ref{Elimits}) as the limits of the integrals of Eq. (\ref{ch3-48}), expressed in terms of the dimensionless parameters.
As in the non-relativistic case, in order to grow a point mass adiabatically within the Hernquist model,
the argument $\te'$  of the initial distribution function becomes a function of $\te$ and $\tl$ by 
equating the radial actions and using the fact that $\tl'=\tl$ from the angular action. Hence, the density around a relativistic point mass in a Hernquist profile takes the form:
\be \label{ch3-51}
\rho(r)=\frac{1}{ \sqrt{2} (2\pi)^2 } \left (\frac{M}{a^3} \right )\frac{a}{r-2Gm} \int_0^{\te_{\rm max}} [1-(GM/a)\te] \ d \te \int_{\tls_{\rm min}}^{\tls_{\rm max}} d \tls \  \frac{\tilde{f}_H\left (\te' (\te,\tl ) \right)}{\sqrt{\tls_{\rm max}-\tls}} \ ,
\ee
The difference here is that in equating the radial actions in Eq. (\ref{eq:adiabatic}), we use the relativistic expression for the point-like mass radial action i.e. Eq. (\ref{Ir_rel}) which in terms of dimensionless variables can be written as 
\be \label{ch3-52}
I_{\rm {r, \ rel}}^{\rm bh}=2\sqrt{GMa}\int_{x_-}^{x_+}  \left [2(\tm/x-\te)-\tls/x^2+\te^2 \ GM/a+(2GM/a) (\tm/x) (\tls/x^2) \right ]^{1/2} \, dx \ , 
\ee
where $x_+$ and $x_-$ are the two turning points. The integration in Eq.~ (\ref{ch3-52}) will have to be evaluated numerically. Now we take the same steps as we used to get Eq. (\ref{Ir_H}): first we combine the terms inside the square root to get
\begin{eqnarray} \nonumber
2(\tm/x-\te)-\tls/x^2+\te^2 \ GM/a+(2GM/a) (\tm/x) (\tls/x^2) \\ \label{ch3-53}
=\frac{-2\te(1-\te \ GM/2a)x^3+2\tm x^2-\tls x+2\tm \tls \ GM/a}{x^3} \ .
\end{eqnarray}
We solve for the three roots of the numerator, of which the two positive roots give the turning points $x_+$ and $x_-$, while the third $x_{\rm neg}$ is always negative. We then rewrite the function in the square root as:
\be \label{ch3-54}
2\te(1-\te \ GM/2a)\frac{(x_+-x)(x-x_-)(x-x_{\rm neg})}{x^3}
\ee
which is positive in the region $x_- \le x \le x_+$. We now make a change of variables $x = t \left(x_+ - x_-\right) + x_-$, which brings the integral into the domain $[0,1]$:
\be \label{ch3-55}
I_{\rm {r, \ rel}}^{\rm bh}=2\sqrt{GMa} \sqrt{2\te(1-\te \ GM/2a)} (x_+-x_-)^2 \int_0^1dt\sqrt{\frac{(x_+-x)(x-x_-)(x-x_{\rm neg})}{x^3}}
\ee
As before, this leads to easier numerical control.

For $\tls=0$, the radial invariant can be integrated analytically, with the turning points $x_-=0$ and $x_+=\tm/(\te(1-\te \ GM/2a))$:
\begin{eqnarray}  \nonumber 
I_{\rm {r, \ rel}}^{\rm bh}&=&2 \sqrt{GMa}\int_0^{\tm/(\te(1-\te \ GM/2a))} d x \sqrt{2\left(\frac{\tm}{x}-\te \right)+\te^2 \frac{GM}{a}} \ ,  \\  \label{ch3-56}
&=&2 \pi \sqrt{GMa}\frac{\tm}{\sqrt{2 \te}\sqrt{1- \te \ GM/2a}} \ .
\end{eqnarray}
and we use this fact in the code. The radial invariant  is again divergent for $\epsilon \rightarrow 0$ but we are only interested in finding a solution in the domain $(0,1]$. For the Hernquist potential we use the same equations as the non-relativistic calculations.

\begin{figure}
\begin{center}
\includegraphics[width=4in]{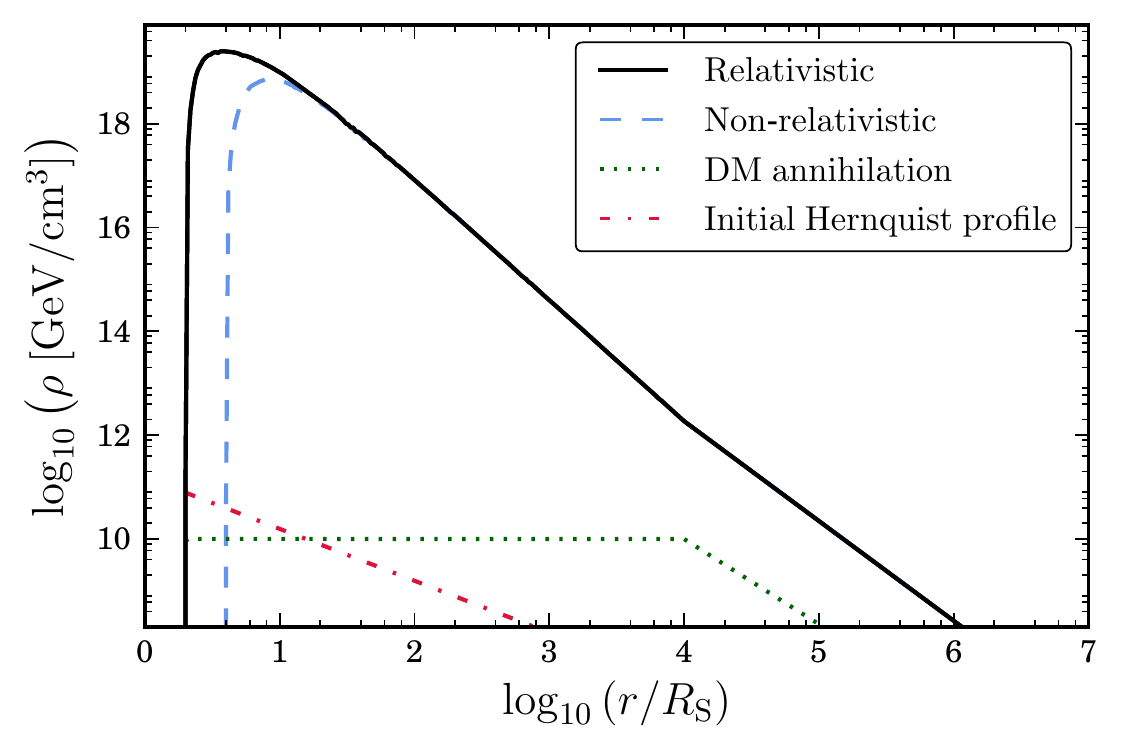}
\caption{Effect of the adiabatic growth of the supermassive black hole at the center of the galaxy on a Hernquist dark matter profile. Shown are the results of the fully relativistic calculation, and the effects of dark matter annihilations. The dashed line (blue in color version) shows the non-relativistic approximation using the GS method.}
\label{fig:hernquist}
\end{center}
\end{figure}

Again we remap the integral in Eq. (\ref{ch3-51}) into a square domain using Duffy transformations. The only difference here is that $\tl^2_{\rm min}$ also depends on $u$. With these changes, the integral in Eq. (\ref{ch3-51}) reads:
\begin{eqnarray}
 \label{ch3-50}
\rho(r) &=&\frac{1}{ \sqrt{2} (2\pi)^2 }  \left (\frac{M}{a^3} \right )\frac{ a \te_{\rm max}}{r-2Gm} 
\nonumber \\
&& \qquad \times \int_0^1 du \int_0^1 d z \ [1-(GM/a)\te_{\rm max} u] \sqrt{\frac{\tls_{\rm max}(u)-\tls_{\rm min}(u)}{1-z}} \tilde{f}_H \left (\te'(u,z)\right) \ ,
\end{eqnarray}
where the arguments of the distribution function are given in Eq. (\ref{eq:uz}). 
The numerical integrations yield the curve labeled ``Relativistic'' in Fig.\ \ref{fig:hernquist}.

\subsection{Profile Modification due to self-annihilation} 

Our calculations so far give the dark matter distribution as it reacts to the gravitational field of the growing black hole. In addition, the dark matter density will decrease if the particles self-annihilate. In fact, if we take into account the annihilation of dark matter particles, the density cannot grow to arbitrary high values, the maximal density being fixed by the value  \cite{Bertone05}:
%To estimate this effect, write a quantity with units of density out of the relevant quantities:
\begin{equation}
 \label{eq:rhocore}
\rho_{\rm{core}}=\frac{m_\chi}{\sigma v \: t_{\rm{bh}}},
\end{equation}
where $\sigma v$ is the annihilation flux, $m_\chi$ is the mass of the dark matter particle, and $t_{\rm{bh}}$ is the time over which the annihilation process has been acting, which we take to be $\approx 10^{10}$yr \cite{GS}.

The probability for dark matter self-annihilation is proportional to the square of the density,
\be
\dot{\rho}=-\sigma v \frac{\rho^2}{m_\chi} = 
- \frac{\rho^2}{\rho_{\mathrm{core}}{t_\mathrm{bh}}}.
\label{eq:rhodotcore}
\ee
This expression can be derived by noting that the annihilation rate per particle is $\Gamma = n \sigma v$, therefore $\dot n=-n \Gamma=-n^2\sigma v$ and $\rho = n m_\chi$.

If we call the output of our code neglecting self-annihilations $\rho'(r)$ 
and the final spike profile  $\rho_\mathrm{sp}(r)$,
we can integrate Eq.~(\ref{eq:rhodotcore}) as follows:
\be
\int_{\rho'(r)}^{\rho_\mathrm{sp}(r)}{\frac{\rho_\mathrm{core} \: d
\rho}{\rho^2}} = - \int_0^{t_\mathrm{bh}}{\frac{d t}{t_\mathrm{bh}}},
\ee
which gives:
\be
\rho_{\mathrm{sp}} (r) = \frac{\rho_\mathrm{core} \rho'(r)}
{\rho_\mathrm{core}+ \rho'(r)}.
\label{rhospike}
\ee

Our calculations do not include the effect of the gravitational field of the halo in the final configuration. This is a good approximation close to the black hole, but far away from the center the effect of the black hole is negligible and the dark matter density will be described by the halo only.  We take care of this fact by simply adding the initial Hernquist profile, given in Eq. (\ref{hernquist0}), to the calculated spike (\ref{rhospike}). We expect this approximation to be good, except possibly in the transition region.
The result is the curve labeled ``DM annihilation'' in Fig.~\ref{fig:hernquist},
where we chose numerical values adequate for a weak-scale thermal relic,
$\sigma v = 3 \times 10^{26} \mathrm{cm^3/s}$ and $m_\xi = 100$ GeV. Note that
whenever an annihilation core is formed,
we can take into account the general relativistic corrections simply by 
allowing the core to extend down to $4Gm$, instead of $8Gm$.

%-------------------------------------------
\section{Pericenter Precession with a Dark Matter Spike}
\label{sec:pericenter}

\begin{figure}
\begin{center}
\includegraphics[width=4in]{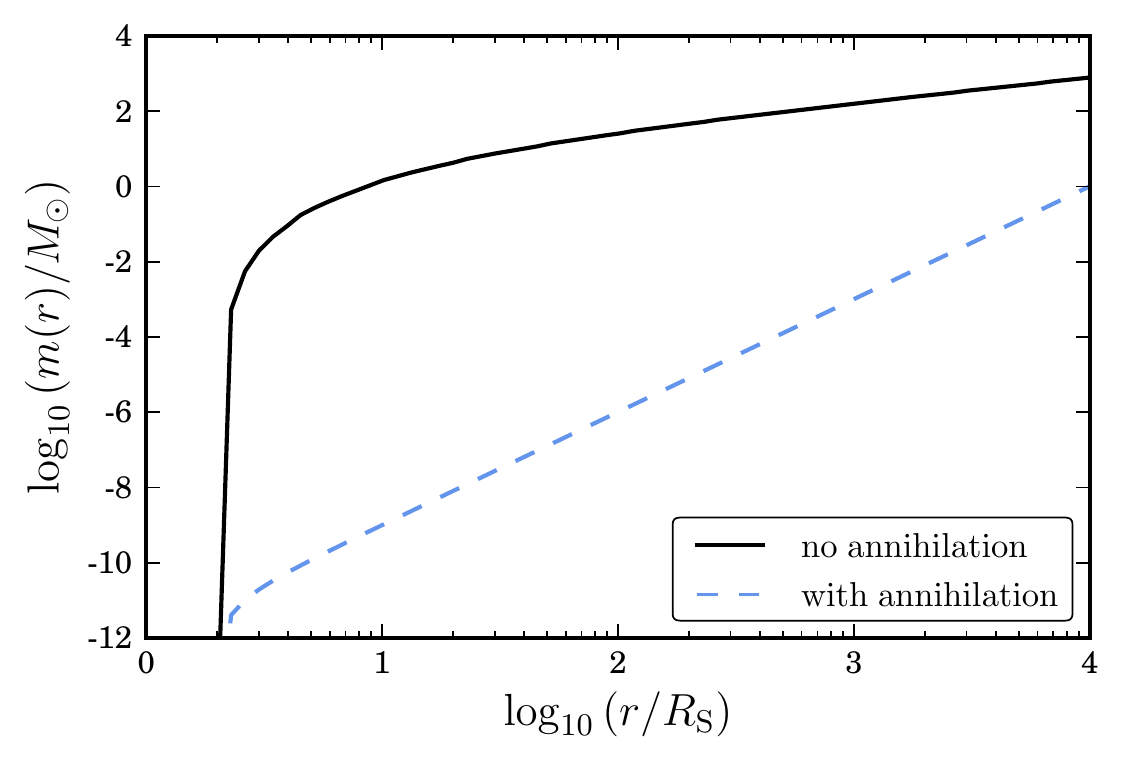}
\caption{Dark matter mass inside radius $r$ for the non-self-annihilating and self-annihilating cases. }
\label{mr}
\end{center}
\end{figure}

The presence of the dark matter density at the galactic center can perturb the orbits of stars in that region.   A spherically symmetric distribution of dark matter will cause pericenter precessions in orbital motions, but will
not change the orientation of the orbital planes.  But to get an upper bound on the possible effect of a non-spherical distribution of dark matter on the orbits of potential no-hair-theorem target stars, it is useful to determine the pericenter precession.  For this we need the dark matter mass inside a given radius $r$, which we obtain by
integrating our density profiles.  The result, for both the self-annihilating and non-self-annihilating cases, is shown in Fig.~\ref{mr}.
As can be seen from Fig.~\ref{mr}, we can approximate the total mass of the dark matter in the region between $10 \,R_{\rm S}$ and $10^4 \, R_{\rm S}$ by a power-law function:
\be \label{ch3-58}
m(r)=m_0({r}/{r_0})^q \ ,
\ee
which leads to the following additional acceleration term in the equation of motion of a star orbiting the black hole:
\be \label{ch3-59}
\bm A=-G\frac{m(r)}{r^2} \ \hat{\bm n}=-\frac{Gm_0 r^{q-2}}{r_0^q} \ \hat{\bm n}\ ,
\ee 
where $\hat{\bm n}\equiv {\bm r}/r$.  Using the standard theory of perturbation of orbital elements, we find that the rate of change with angle of the pericenter of an orbit is given by
\be \label{ch3-60}
\frac{d \omega}{df}=\frac{r^2}{h} \frac{d \omega}{dt}=-\frac{r^2 p}{e h^2} \left ({\bm A} \cdot \hat{\bm n} \right ) \cos f  \,,
\ee
where $h = \sqrt{Gmp}$ is the angular momentum per unit mass, $p = a(1-e^2)$ is the semi-latus rectum, $a$ is the semi-major axis (not the Hernquist scale, nor the Kerr parameter), $e$ is the eccentricity, and $f = \phi -\omega$ is the true anomaly (angle from pericenter).
 Substituting Eq. (\ref{ch3-59}) in Eq. (\ref{ch3-60}) and using $r=p/(1+e\cos f)$, we get
\be 
\label{ch3-61}
\frac{d \omega_{\rm DM}}{df}=\frac{1}{e}\left( \frac{m_0}{m} \right) \left( \frac{p}{r_0} \right)^q \frac{\cos f}{(1+e \cos f)^q} \ .
\ee
To get the change of $\omega$ over one orbit, we integrate Eq. (\ref{ch3-61}) over the true anomaly $f$ from $0$ to $2\pi$ to obtain
\begin{eqnarray} 
\Delta \omega_{\rm DM}  
&=&- \pi q \left( \frac{m_0}{m} \right) \left( \frac{p}{r_0} \right)^q (1-e^2)^{1/2} f_{\rm q}(e) \ ,
\label{ch3-62}
\end{eqnarray}
where, for various values of $q$, we get the forms for $f_{\rm q}(e)$ shown in Table \ref{table:f}

\begin{table}
\begin{center}
\begin{tabular}{ccccc}
\hline
%\multicolumn{2}{c}{Item} \\
%\cline{1-2}
{\rm q}    && $f_{\rm q}(e)$  && Range of $f_{\rm q}(e)$  \\
\hline\hline
1      && $2(1+\sqrt{1-e^2})^{-1}$    && $[1,2]$      \\
2      && 1     && 1      \\
3      && 1     && 1      \\
4      && $1+e^2/4$   && $[1,5/4]$    \\
\hline
\end{tabular}
\caption{The function $f_q(e)$.}
\label{table:f}
\end{center}
\end{table}

Now from Fig.~\ref{mr}, we can see that the power $q$ in Eq. (\ref{ch3-58}) can be chosen to be $3$ or $1$ depending on whether the dark matter particles self annihilate or not, respectively. Using $r_0=R_{\rm S}\times 10^4 \approx 4.6 \ {\rm mpc}$, assuming a black-hole mass $m=4 \times 10^6 M_\odot$, we can read off the values of $m_0$:
\be  \label{ch3-63}
m_0=\left\{ \begin{array}{ll}
    10^3 \ M_\odot \,, &q=1 \quad \text{no self-annihilation}\,,\\
    1 \ M_\odot \,, &q=3 \quad \text{self-annihilation} \,.
     \end{array} \right.
\ee

An estimate of the  astrometric effect $\dot{\Theta}$ of the pericenter precession as seen from Earth is given by the rate of precession at the source $\Delta \omega/P$, where $P=2\pi (a^3/Gm)^{1/2}$ is the orbital period, multiplied by $a/D$, where $D$ is the distance to the galactic center. Using Eqs.~(\ref{ch3-62}) and (\ref{ch3-63}), together with $D=8 \ {\rm kpc}$, we obtain the rates for the non-self-annihilating ($q=1$) and self-annihilating ($q=3$) cases in microarcseconds per year:
\begin {eqnarray}
\dot \Theta_{\rm DM, no-ann}&=&6.26 \ P^{1/3} \frac{\sqrt{1-e^2}}{1+\sqrt{1-e^2}}  \;\;\; \mu {\rm arcsec/yr} \ , \\
\dot \Theta_{\rm DM, ann}&=&3.81 \times 10^{-4} \ P^{5/3} \sqrt{1-e^2}  \;\;\; \mu {\rm arcsec/yr} \ .
\end{eqnarray}

To compare the rate of precession of pericenter of a star rotating the black hole induced by dark matter with the relativistic effects of the black hole, we provide in Table~\ref{omega-DM},  numerical results for the S2 star and for a hypothetical target star which is closer to the center and could be used for the test of the no-hair theorem. Shown are the astrometric pericenter precessions rates as seen from Earth from the Schwarzschild part of the metric and from the two dark matter distributions ($\dot \Theta_{\rm S}$, $\dot \Theta_{{\rm DM,ann}}$, and $\dot \Theta_{{\rm DM,no-ann.}}$, respectively) and the orbital plane precessions from the frame dragging and quadrupole effects of the black hole, $\dot \Theta_{\rm J}$ and $ \dot \Theta_{{\rm Q}}$, respectively (see \cite{cwnohair} for the relevant formulae for the three relativistic effects).
\begin{table}
\begin{center}
\begin{tabular}{lcccc}
\hline
&& S2 Star  && No-hair Star  \\
\hline\hline
$a$ (mpc)&& 4.78 && 0.2  \\ 
$e$&&0.88&&0.95 \\
$P$ (yr)&&15.5 &&0.13 \\
\hline
{$\dot \Theta_{\rm S}$} && 26.533    && 7319.92      \\
{$\dot \Theta_{\rm J}$} && 0.235     && 486.303      \\
{$\dot \Theta_{{\rm Q}}$} && 0.002     && 36.325      \\
{$\dot \Theta_{{\rm DM,no-ann.}}$} && 5.026   && 0.755    \\
{$\dot \Theta_{{\rm DM,ann.}}$} && 0.017   && $4 \times 10^{-6}$  \\
\hline
\end{tabular}
\caption{Astrometric precession rates as seen from the Earth in units of $\mu$arcsec/yr; $\dot \Theta_{\rm J}$ and $\dot \Theta_{{\rm Q}}$ denote orbital plane precessions, while the others denote pericenter precessions}
\label{omega-DM}
\end{center}
\end{table}

\begin{figure}
\begin{center}
\includegraphics[width=4in]{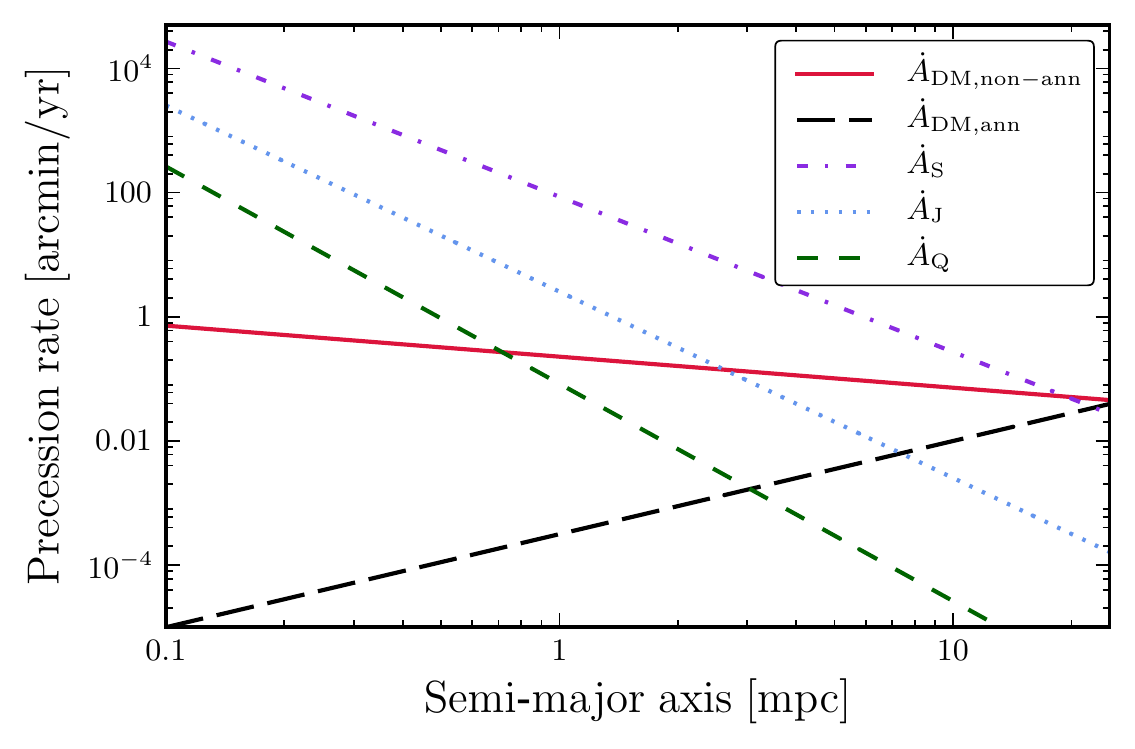}
\caption{Precession rates  at the source for  a star with $e=0.95$ induced by relativistic effects of the central black hole and by distributions of non-self-annihilating and self-annihilating dark matter. Shown are the pericenter precession rates from relativistic (dot-dashed/purple) and dark matter (solid/red; long-dashed/black) effects, and the orbital plane precession rates from relativistic frame dragging (dotted/blue) and quadrupolar (dashed/green) effects.}
\label{precession_rates}
\end{center}
\end{figure}

As an alternative way to compare the orbital precessions induced by dark matter and by the black hole, we compute the amplitudes of various precession rates as seen at the source.   From Eq.\ (3) of \cite{cwnohair} and from the results of this section, we list the precession rate amplitudes: 

\begin{eqnarray}
{\dot A}_S &=&\frac{6\pi}{P} \frac{Gm}{a(1-e^2)} 
 \approx  8.335 \ \tilde a^{-5/2} (1-e^2)^{-1} \ {\rm arcmin/yr} \ , 
 \nonumber \\
\dot{A}_J&=& \frac{4\pi}{P} \chi \left[ \frac{Gm}{a(1-e^2)}\right]^{3/2}  
 \approx 0.0768 \ \chi {\tilde a}^{-3} (1-e^2)^{-3/2} \ {\rm arcmin/yr} \ , 
 \nonumber\\ 
\dot{A}_{Q}&=& \frac{3 \pi}{P} \chi^2  \left[ \frac{Gm}{a(1-e^2)}\right]^2  
 \approx  7.9 \times 10^{-4} \chi^2 {\tilde a}^{-7/2} (1-e^2)^{-2} \ {\rm arcmin/yr} \ , 
 \nonumber\\
%\\ \nonumber
{\dot A}_{\rm DM,  no-ann} &=&\frac{\Delta \omega_{\rm DM, no-ann}}{P}
\approx   0.953 \ {\tilde a}^{-1/2}(1-e^2)^{1/2}[1+(1-e^2)]^{-1/2} \ {\rm arcmin/yr}   \ , 
\nonumber\\
%\\ \nonumber
{\dot A}_{\rm DM, ann}&=&\frac{\Delta \omega_{\rm DM,ann}}{P} \approx  9.8 \times 10^{-5} \ {\tilde a}^{3/2} (1-e^2)^{1/2} \ {\rm arcmin/yr}  \ ,
\end{eqnarray} 
where $\tilde{a}$ is the semi-major axis in mpc, and $0 \le \chi \le 1$ is the dimensionless spin parameter of the black hole.  Figure \ref{precession_rates} plots these amplitudes  for $e = 0.95$, $\chi = 1$, for semi-major axes ranging from $0.1$ to $20$ mpc.

As can be seen from Table~\ref{omega-DM} and Fig.~\ref{precession_rates}, for hypothetical target stars in eccentric orbits with semi-major axes less than about $0.2$ milliparsec, which could be used to test the no-hair theorem, the pericenter precessions induced by the dark matter distribution at the center are much smaller than the relativistic precessions. Because the pericenter advance due the dark matter distribution is so small , we argue that it is reasonable to consider this as a good estimate for the upper limit on the precession of orbital planes that might be induced by a non-spherical component of the dark matter distribution that would be generated by a rotating central black hole. That non-spherical part is likely to be a small perturbation of the basic dark matter distribution because the effects of frame dragging and the quadrupole moment are relativistic effects that  fall off faster with distance than the basic Newtonian gravity of the hole.  In addition the mass of dark-matter inside a relevant orbit is a tiny fraction ($< 10^{-3}$) of the black hole mass, and therefore will not modify the mass inferred from the orbits of stars such as $S2$.
As a result, we can conclude that a dark matter distribution near the black hole will not significantly interfere with a test of the black hole no-hair theorem. Furthermore, if the dark matter particles are self-annihilating, their effects will be utterly negligible.

On the other hand, for S2-type stars, if future observational capabilities reach the level of 10 $\mu$arcsec per year, the perturbing effect of the dark matter distribution on stellar motion at the GC could be marginally detectable if the dark matter particles are not self-annihilating, as would be the case if they were axions, for example. If they are self-annihilating, the effects of a dark matter distribution on the outer cluster of stars will be unobservable.  
For other discussions of the effects of dark matter on stellar orbits see \cite{Zakharov2007,Iorio2013}.

\section{Concluding remarks}
\label{sec:conclusion}

We have carried out a fully relativistic calculation of the effect of an adiabatically grown black hole on the distribution of dark matter at the galactic center, thereby putting the work of Gondolo and Silk \cite{GS} on a firm relativistic footing.   The differences we find are dramatic only in the innermost region, where we find that the density of dark matter extends all the way to $r=4Gm = 2R_{\rm S}$, instead of vanishing at $4R_{\rm S}$.  

Hence, when estimating the effects of the dark matter from the galactic center,
around a population of Intermediate Mass Black Holes~\cite{Bertone:2009kj}
(which could be ubiquitous in the Milky Way halo~\cite{Rashkov:2013uua}), or
at the center of Active Galactic Nuclei~\cite{Gomez:2013qra}, one should keep
in mind the enhanced density of dark matter closer to the black hole.

On the other hand the total amount of mass represented by this innermost region is very small, and therefore the additional perturbing effects on the orbits of stars in the central cluster will be small.    For the same reason the additional density in this innermost region will have a small effect on line-of-sight integrals that give net fluxes of high-energy radiation to be expected either from dark matter decays or annihilations.

Finally, we note that we have adopted the GS adiabatic growth model, cognizant of its limitations.   For the evolution to be adiabatic, the dynamical timescale
inside the region where the black hole dominates should be much shorter than
both the typical timescale for black hole growth and the relaxation timescale
of the dark matter halo~\cite{Sigurdsson:2003wu}. 
The former can be estimated as $t_{dyn} = r_h/\sigma
\sim 10^4$ yr, where $r_h = G m/\sigma^2$ is the region where the black hole
dominates and $\sigma$ is the velocity dispersion of the dark matter
particles. Assuming Edddington accretion, it would take 
$t_S = m/\dot{m}_\mathrm{Edd} \sim 5 \times 10^7$ yr, while for collisionless
dark matter the relaxation timescale is, indeed, longer than $t_{dyn}$. 
Let us note that several effects could invalidate these arguments.
For instance, if the seed black hole is initially off-center~\cite{Ullio:2001fb}, 
if there are hierarchical mergers~\cite{Merritt:2002vj}, or if there is kinetic heating caused
by scattering of the dark matter particles by stars in the dense stellar
cusp around the hole~\cite{Bertone:2005hw}, the GS spike could be destroyed.
Important as they are, however, these effects are unrelated to our main purpose
of understanding the general relativistic effects close to the black hole within the GS model.

\acknowledgments

This work was supported in part by the National Science Foundation,
Grant Nos.\ PHY 09--65133, 12--60995 \& 0855580 and by the U.S. DOE under
contract No. DE-FG02-91ER40628.  We thank the Institut d'Astrophysique de Paris for
its hospitality during part of this work.

%\appendix

%#################

%########  

\end{document}